\documentclass[default,iicol]{sn-jnl}

\jyear{2021}%

\theoremstyle{thmstyleone}%
\theoremstyle{thmstyletwo}%

\theoremstyle{thmstylethree}%

\begin{document}

\title[Article Title]{Impact of Baryon anti-Baryon annihilation on hyperon ($\Lambda$, $\bar\Lambda$) production and apparent strangeness enhancement in $\bar\Lambda/\bar{p}$ in heavy ion collisions at SPS energy}


\author*[1,2]{\fnm{Ekata} \sur{Nandy}} \email{ekata@vecc.gov.in}

\author[1,2]{\fnm{Subhasis} \sur{Chattopadhyay}} \email{sub@vecc.gov.in}


\affil[1]{\orgdiv{Variable Energy Cyclotron Centre}, \orgaddress{\street{1/AF, Bidhan Nagar}, \city{Kolkata}, \postcode{700064}, \country{India}}}

\affil[2]{\orgdiv{Homi Bhabha National Institute}, \orgaddress{\city{Mumbai}, \postcode{400094}, \country{India}}}

\abstract{A deconfined medium of quarks and gluon, called the Quark-Gluon Plasma (QGP) is produced when heavy-nuclei are collided at relativistic energies. The QGP formation is often characterized by a phenomenon called strangeness enhancement where, the relative production of strange-to-non-strange particles are enhanced in central collisions compared to peripheral or proton-proton interactions. Besides the enhancement in K/$\pi$ ratios, a non-monotonic energy dependence was also reported for $\bar{\Lambda}$ to $\bar{p}$ ratios at CERN SPS, attributed to a signature for the strangeness enhancement as well. As anti-particles are produced directly from the reaction, the $\bar{\Lambda}$/$\bar{p}$ ratios are considered as a cleaner probe for the strangeness enhancement. However, at this energy range hadronic interactions have a dominant role to play and, importantly for $\bar{\Lambda}$ and $\bar{p}$, processes like baryon-anti-baryon ($\mathrm{B\bar{B}}$) annihilation can have a significant impact. In this work, we use a hadronic transport model UrQMD, to investigate the role of baryon-anti-baryon ($\mathrm{B\bar{B}}$) annihilation on
$\Lambda$, $\bar{\Lambda}$ hyperon production and its effect on $\bar{\Lambda}$/$\bar{p}$ ratios. The UrQMD calculations that include $\mathrm{B\bar{B}}$ annihilation can produce the trend of average transverse mass spectra for $\Lambda$ and $\bar{\Lambda}$, as well as, the characteristic enhancement in $\bar{\Lambda}$/$\bar{p}$ ratios in data as a function of centrality and collision energy. Furthermore, $\bar{\Lambda}$/$\bar{p}$ ratios extracted from the feed-down corrected SPS data are seen to be in good agreement with UrQMD model calculations with $\mathrm{B\bar{B}}$ annihilation. This suggests that $\bar{\Lambda}$/$\bar{p}$ enhancement is not necessarily because of strangeness enhancement and $\mathrm{B\bar{B}}$ annihilation has a significant role to play. }




\maketitle

\section{Introduction}\label{sec1}
Quantum Chromodynamics (QCD) based on lattice formalism have predicted that a deconfinement phase transition may occur when heavy-nuclei (A+A) are collided at relativistic energies~\cite{lat_Karsh,lat_Cheng}. Such a phase transition is likely to produce a novel form of QCD matter consisting of a deconfined state of quarks and gluons (partons), the Quark Gluon Plasma (QGP). Over the last two decades, data from the Relativistic Heavy Ion Collider (RHIC) in BNL and CERNs' Large Hadron Collider (LHC) have presented convincing evidences in favour of this partonic deconfinement ~\cite{BRHAMS_NPA,PHOBOS_NPA,PHENIX_NPA,STAR_NPA, ALICE_ARNPS}.  

For a long, strangeness enhancement is considered as one of the most important signatures for the QGP formation~\cite{JR_BMu_PRL_48_1066}. Strangeness production was proposed to be enhanced in a partonic environment because of abundant production of strange-antistrange ($\mathrm{s\bar{s}}$) quark pairs from gluon fusion. These strange quarks/antiquarks may eventually coalesce with other non-strange or strange (anti)quarks from the deconfined partonic medium to produce strange hadrons~\cite{Reco_Coalescence_1, Reco_Coalescence_2, Reco_Coalescence_3, Reco_Coalescence_4,  Reco_Coalescence_5, Reco_Coalescence_6}. Enhancement in the yields of strange particles and/or of the ratios of yields of strange-to-non-strange particles have been reported in central heavy-ion collisions when compared with elementary pp interactions~\cite{MM_NA49_JPG_43}. Nevertheless, this relative strangeness enhancement from a small to large system can also occur in a pure hadronic environment because of the volume effect. Due to the requirement of local strangeness (quantum number) conservation, strangeness production in small systems is canonically suppressed~\cite{Canon_SE_SH_PLB, AT_KR_JPG28_2092}. This constrain is however, relaxed for a large system where strangeness neutrality may be achieved globally~\cite{AT_KR_JPG28_2092}. Thus, it reduces the phase space penalty for the production of a strange or multi-strange hadrons in A+A collision as the strangeness conservation can be taken care-of by another anti-(multi-)strange hadron elsewhere in the extended volume. Consequently, even in the absence of partonic medium, ratios of yields of strange particles in A+A over pp collisions can exhibit an enhancement simply because the strangeness production in small systems are canonically suppressed.

Strangeness enhancement has been extensively studied at RHIC-AGS and CERN-SPS. A non-monotonic variation in K$^{+}$/$\pi^{+}$ ratios as a function of beam energy, popularly known as the horn structure, was reported~\cite{SE_diff_collab_1, SE_diff_collab_2, SE_diff_collab_3, STAR_PRC_2017}. This non-trivial energy dependence in the K$^{+}$/$\pi^{+}$ ratios was speculated as a signature of strangeness enhancement in 
high energy heavy-ion collisions however, there also exists some alternative explanations~\cite{CSR_PHSD}. Similar signatures of enhancement were also observed in the strange baryon sector where prominent increase in the yields of $\Lambda$-hyperons and multi-strange hyperons $\Xi$ and $\Omega$ have been demonstrated~\cite{Antinori_wa97_epjc, Antinori_na35_jpg, CAlt_na49_prl94, AL_by_Ap_HZ, CAlt_NA49_nucl_0804}. In terms of strange-to-non-strange particle ratios, a large enhancement in $\bar{\Lambda}/\bar{p}$ ratio from 0.26 in peripheral collisions to $\sim$3.6 in central Au+Au collisions at 11.7 AGeV was reported by the E917 experiment at AGS~\cite{E917_PRL_87_242301}, although with large uncertainty. A large value of $\bar{\Lambda}/\bar{p}$ ratio of 2.9 $\pm$ 0.9(stat) $\pm$ 0.5(sys) in central Si+Au collisions at 14.6 AGeV was also measured by the E859 experiment~\cite{E859}.

Considering inclusive $\bar{p}$ yields from two different experiments- E864 and E878,
an indirect estimate of $\bar{\Lambda}/\bar{p} > $ 2.3 was obtained in central Au+Pb collisions at 11.5 AGeV at y$\sim$ 1.6 and p$_{T}\sim$0~\cite{E864}. Also, a significant rise from 0.25
in pp collisions to 1.5 for heavy-ion systems was published by the NA35 experiment at CERN SPS \cite{JG_NA35_NPA}. Subsequent studies in Pb+Pb collisions between 20 to 158 AGeV by NA49 experiment supports earlier findings of large enhancement in $\bar{\Lambda}/\bar{p}$ ratios \cite{CAlt_NA49_nucl_ex0512033}. In terms of quark composition, $\bar{\Lambda}$ and $\bar{\mathrm{p}}$ are unique because they are comprised of anti-quarks alone, which can only be produced during reaction. As in the entrance channel of nuclear collisions there are no antibaryons or strangeness, it might appear that this large enhancement in $\bar{\Lambda}/\bar{p}$ ratios is because of  partonic deconfinement~\cite{PK_BM_JR_Phys_Rept}. However, it must be remembered, at this concerned energy range, high baryon densities may have a significant impact on the final spectra and yields of strange and ordinary baryons or anti-baryons, of particular interest, for $\bar{\Lambda}$ and $\bar{p}$ is baryon-antibaryon($\mathrm{B\bar{B}}$) 
annihilation. Importantly, the $\mathrm{B\bar{B}}$ annihilation effects are shown to be more effective at 
SPS energies~\cite{ES_phsd_bbbar_dyamics,FW_PRC_2012}.

It is known that $\bar{\Lambda}$ and $\bar{p}$ can have different annihilation cross-sections and it may be possible that because of this difference final yields of $\bar{\Lambda}$ and $\bar{p}$ are modified differently, which may cause an apparent enhancement in $\bar{\Lambda}$/$\bar{p}$ ratios. Therefore, before attributing $\bar{\Lambda}/\bar{p}$ enhancement to a signature for strangeness enhancement and hence to the formation of QGP, it is crucial that we understand the impact of $\mathrm{B\bar{B}}$-annihilation
on $\bar{\Lambda}$ and $\bar{p}$ production in detail.
  
In this work, our goal is to demonstrate the sensitivity of ${\Lambda}$-hyperon production
and $\bar{\Lambda}/\bar{p}$ ratio to $\mathrm{B\bar{B}}$-annihilation by analyzing average transverse mass of ${\Lambda}$ $\&$ $\bar{\Lambda}$ and ratios of $\bar{\Lambda}$-to-$\bar{p}$  as a function of centrality and collision energy in central Pb+Pb collisions at $\sqrt{s_{NN}} =$ 6.27, 7.62, 8.77, 12.3 and 17.3 GeV (corresponding to the E$_{lab}$= 20, 30, 40, 80 and 158 AGeV, respectively) with and without including the effect of $\mathrm{B\bar{B}}$ annihilations in a hadronic transport model, UrQMD (Ultra-Relativistic Quantum Molecular Dynamics)
~\cite{UrQMD_MB_JPG_25_1859, UrQMD_SB_PPNP_41_255}. As the dynamics of particle production at SPS energies are believed to be dominated by hadronic interactions so we choose UrQMD model because it describes the phenomenology of particle production at SPS energy reasonably well.

The organization of this paper is as follows. In section 2 we briefly describe the UrQMD transport model and implementation of $\mathrm{B\bar{B}}$-annihilation in UrQMD followed by results and discussions in section 3. Finally we summarize our findings in section 4

\section{UrQMD}\label{sec2}
UrQMD (Ultra-relativistic Quantum Molecular Dynamics) is a microscopic transport model that provides a unified framework to describe the phenomenology of particle production in pp, p+A and A+A collisions over a broad energy range \cite{UrQMD_SB_PPNP_41_255}.
This model is based on an effective solution to relativistic Boltzmann equation on a non-equilibrium approach as shown below,

\begin{equation}
\textit{p}^{\mu}\partial_{\mu}\textit{f}_{i}(\textit{x}^\nu,\textit{p}^\nu) = \textit{C}_{i}.
\end{equation} 

The above equation describes the time evolution of the distribution function $\textit{f}_{i}$ in coordinate and momentum space for a particle species "$\textit{i}$"  and includes the full collision term $\textit{C}_{i}$. The underlying degrees of freedom in UrQMD are hadrons and strings. Here an individual particle propagates on a straight line until the relative distance between two particles is smaller than a critical distance given by the total interaction cross-section between two particles. These cross-sections in UrQMD are either calculated by the principle of detailed balance or additive quark model (AQM) \cite{aqm1,aqm2} or parameterized from the available experimental data. For resonance excitations or decays the Breit-Wigner formalism is implemented. 
In UrQMD, the particle production dynamics is either governed by the decays of baryon or meson resonances or via a string excitation and fragmentation.
Until now, UrQMD has 55 Baryons and 32 Mesons that also include ground state particles and all resonances with mass upto 2.25 GeV. 

In UrQMD, to model $\mathrm{p\bar{p}}$ annihilation cross-section, the elementary annihilation cross-sections are obtained by fitting the available data. It is to be noted that UrQMD uses same data driven parameterization for $\mathrm{p\bar{p}}$ and $\mathrm{B\bar{B}}$ annihilation cross-sections at same $\sqrt{s}$. 
The parameterized annihilation cross-section used in UrQMD is shown below,

\begin{equation}
 \sigma^{\mathrm{p\bar{p}}}_{ann} = \sigma^{\mathrm{N}}_{0} \frac{s_{0}}{s} [ \frac{A^{2}s_{0}}{ (s-s_{0})^{2} + A^{2}s_{0}} + B ]. 
\end{equation}

Where $\sigma^{\mathrm{N}}_{0}$ = 120 mb, $s_{0}$ = 4m$^{2}_{N}$ , A = 50 MeV and B = 0.6 \cite{UrQMD_MB_JPG_25_1859}.

For annihilation channels that involve a strange-baryon/antibaryon, such as $\bar{\Lambda}\mathrm{p}$ or $\Lambda\bar{\mathrm{p}}$, an additional correction factor is introduced based on AQM, given by
\begin{equation}
\sigma^{\mathrm{B\bar{B}}}_{ann} = (1- 0.4\frac{s_{B}}{3} )(1- 0.4\frac{s_{\bar{B}}}{3} ) \sigma^{\mathrm{p\bar{p}}}_{ann} \cite{UrQMD_MB_JPG_25_1859},
\end{equation}
where $s_{B}$ and $s_{\bar{B}}$ is the strangeness number for baryon and antibaryon, respectively.
Thus, annihilation cross-section of $\bar{\Lambda}-\mathrm{p}$,  ($\sigma^{\bar{\Lambda}\mathrm{p}}_{ann}$) is less than annihilation cross-section of p-$\bar{\mathrm{p}}$ ($\sigma^{\mathrm{p}\bar{\mathrm{p}}}_{ann}$).
From equation [2] we see the annihilation cross-section  ($\sigma_{ann}$) has an approximate $\frac{1}{\sqrt{s}}$ dependence with beam energy.

Before proceeding to next section, we would like to draw attention on the fact that UrQMD in it's current implementation
does not include the reverse channels for $\mathrm{B\bar{B}}$ annihilation i.e, the regeneration of baryon-anti-baryon pairs.
This is mainly because of the
limitation in modeling the collision criterion in UrQMD, which is based on the geometric interpretation of the cross section, i.e. 2 particles collide when their distance of closest separation is less or equal to $\sqrt{\sigma /\pi}$, $\sigma$  being the total interaction cross section. This setup does not generalize to collisions involving more than two particles, resulting in the violation of the principle of detailed balance in UrQMD for the processes that have more than 2 particles in the outgoing channel. The detailed balance is however, respected in UrQMD for 2 $\rightarrow$ 2 processes.
 
As the reverse channel for annihilation, i.e, regeneration of 
$\mathrm{B\bar{B}}$ pair is currently missing in UrQMD this might overestimate the effects of the $\mathrm{B\bar{B}}$ annihilation. But, it has to be remembered that for an expanding system, cross section of reverse channel decreases in some powers of phase space density of particles ($\rho$). In particular, for $\mathrm{B\bar{B}}$ annihilation that produces 5$\pi$ on an average in the output channel, a reverse reaction would require scattering of 5$\pi$ $\rightarrow$ $\mathrm{B\bar{B}}$ pair and it's cross section would scale as $\rho^{5}$. As a result, the probability of such reverse reactions fall at a faster rate ~\cite{steinheimer}. Thus, even if the regeneration of $\mathrm{B\bar{B}}$ pair is not implemented in UrQMD, it's impact on the final baryon yields may not be very large ~\cite{PanScott}. 
Nevertheless, for a systematic understanding, 
one can study its effect in some recent models ~\cite{smash} where the principle of detailed balance for $\mathrm{B\bar{B}}$ annihilation is properly incorporated.

\section{Results}\label{sec3}

\begin{figure}[htbp]
		\centering
	\includegraphics[width=80mm,height=60mm]{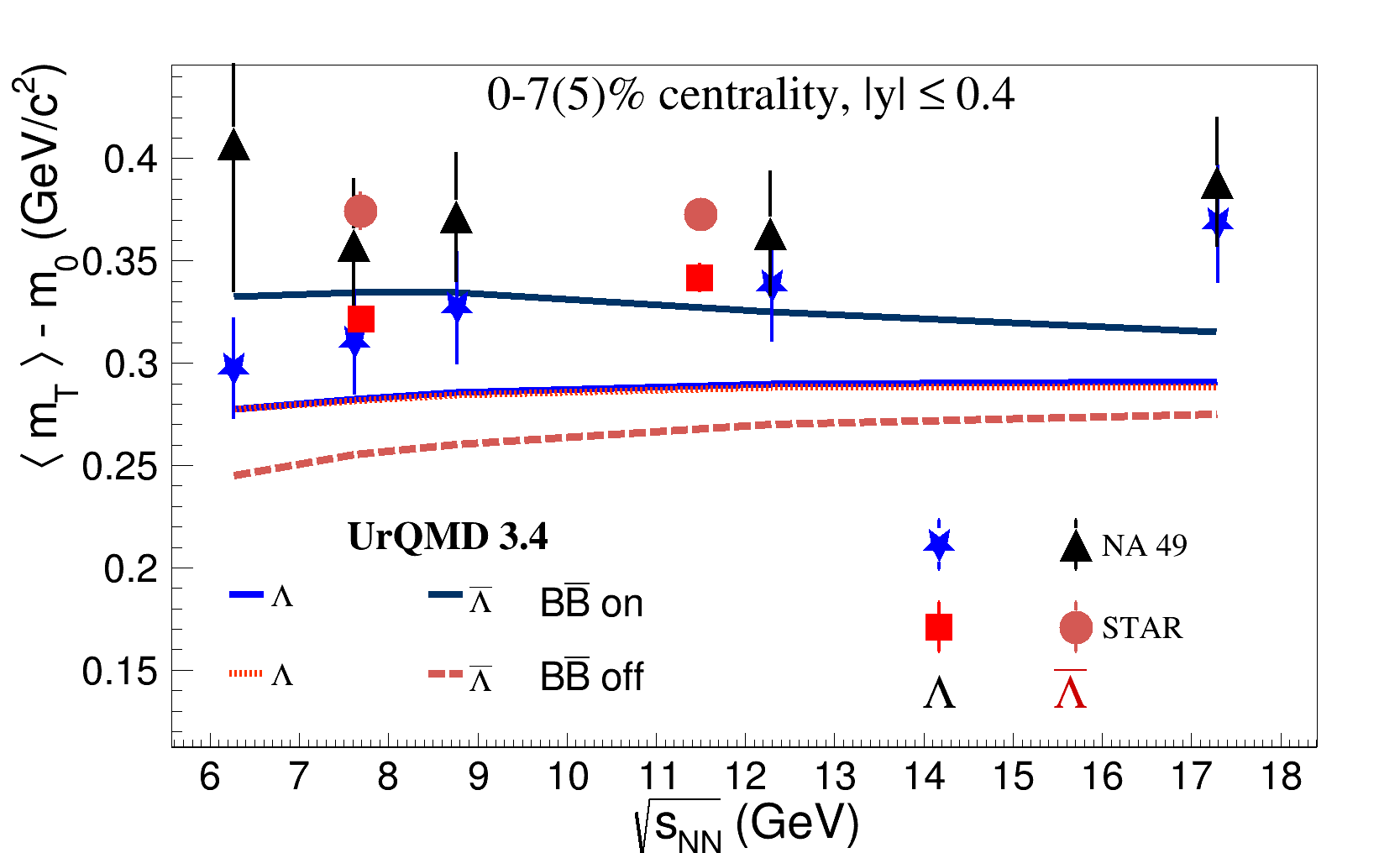}
	\caption{\label{mt_energy}Energy dependence of mean transverse mass, $\langle \mathrm{m_{T}} \rangle$ - m$_{0}$, at
mid-rapidity($\mid$ y $\mid$ $<$ 0.4) for $\Lambda$ and $\bar{\Lambda}$ in central Pb+Pb collisions at $\sqrt{s_{NN}}$ = 6.27 to 17.3 GeV from UrQMD with $\mathrm{B\bar{B}}$ annihilation and without $\mathrm{B\bar{B}}$ annihilation. UrQMD calculations are compared with existing results from central Pb+Pb collisions at $\sqrt{s_{NN}}$ = 6.27 to 17.3 from NA49~\cite{CAlt_NA49_nucl_0804, NA49_data_table} and STAR results from central Au+Au collisions at $\sqrt{s_{NN}}$ = 7.7 and 11.5 GeV~\cite{STAR_PRC_2020}  }
\end{figure}

For this study, we generated about 1 million Pb+Pb UrQMD events with impact parameter, b $<$ 3.4 fm at $\sqrt{s_{NN}}$  = 6.27 GeV, 7.62 GeV, 8.77 GeV, 12.3 GeV and 17.3 GeV in two configurations i.e, with and without incorporating the $\mathrm{B\bar{B}}$ annihilation. As a first step, we study the effect of $\mathrm{B\bar{B}}$ annihilation on the average transverse mass ($\langle m_{T} \rangle$- m$_{0}$) of $\Lambda$ and $\bar{\Lambda}$ as a function of $\sqrt{s}$. 
The average transverse mass represents an effective temperature (T$_{eff}$) of a thermalized medium that include the original thermal energy (T$_{th}$) of the medium and the particle-specific kinetic energy from the hydrodynamic radial flow ($\beta$). The kinetic energy of particles in a thermal medium depend on the mass of the particle alone (m$\beta^{2}$). Massive particles have higher T$_{eff}$ and vice-versa. Interestingly, although $\Lambda$ and $\bar{\Lambda}$ have same mass, their $\langle m_{T} \rangle$- m$_{0}$ are however, different in magnitude and the difference is seen to increase with the decrease in beam energy or increase in net baryon-density~\cite{STAR_PRC_2020}. In Fig.~\ref{mt_energy}, UrQMD calculations for $\langle m_{T} \rangle$- m$_{0}$ for $\Lambda$ and $\bar{\Lambda}$ as a function of $\sqrt{s}$ are shown together with the data points from NA49 and STAR~\cite{CAlt_NA49_nucl_0804, NA49_data_table, STAR_PRC_2020}. UrQMD calculations also show a similar trend in the difference in magnitudes of $\langle m_{T} \rangle$- m$_{0}$ for $\Lambda$ and $\bar{\Lambda}$ as a function of $\sqrt{s}$. This observed split in the magnitude of $\langle m_{T} \rangle$- m$_{0}$ for $\Lambda$ and $\bar{\Lambda}$ can be understood as a consequence of $\mathrm{B\bar{B}}$ annihilation. In a medium with high baryon density, low momenta $\bar{\Lambda}$s have higher chance to annihilate with $\textit{p}$ compared to ${\Lambda}$s that annihilate with $\bar{p}$.
This causes lowering of low p$_{T}$ yields while keeping the high p$_{T}$ yields unchanged, resulting in a hardening of $\bar{\Lambda}$'s p$_{T}$-spectra and leading to systematically higher values of average $\langle m_{T} \rangle$- m$_{0}$ for $\bar{\Lambda}$. This difference in the magnitude of $\langle m_{T} \rangle$- m$_{0}$ between $\Lambda$ and $\bar{\Lambda}$ diminishes with increasing $\sqrt{s}$ as the effect of $\mathrm{B\bar{B}}$ annihilation gradually weakens with decreasing baryon density. Although, UrQMD can qualitatively reproduce the trend in the data well but the quantitative mismatch is because UrQMD underestimates the radial flow.

\begin{figure}[htbp]
		\centering
	\includegraphics[width=82mm,height=48mm]{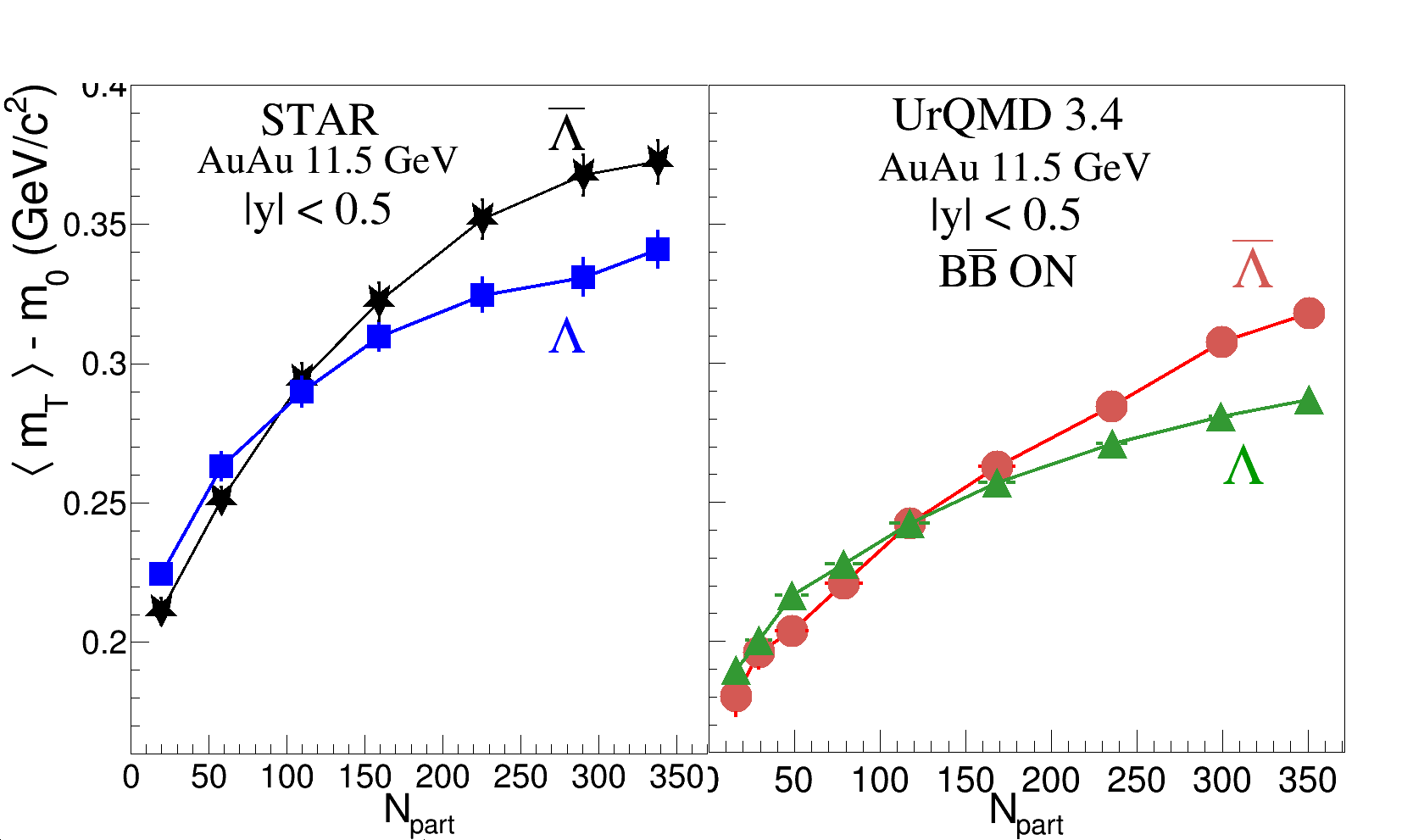}
	\caption{\label{mt_centrality} The $\langle m_{T} \rangle$- m$_{0}$ spectra for ${\Lambda}$ and $\bar{\Lambda}$ as a function of N$_{part}$ as obtained from the STAR data for Au+Au collisions at $\sqrt{s}_{NN}$ = 11.5 GeV (left)~\cite{STAR_PRC_2020} and compared it with UrQMD calculations including $\mathrm{B\bar{B}}$ annihilation at the same $\sqrt{s}_{NN}$ (right) }
\end{figure}

\begin{figure}[htbp]
		\centering
	\includegraphics[width=70mm,height=52mm]{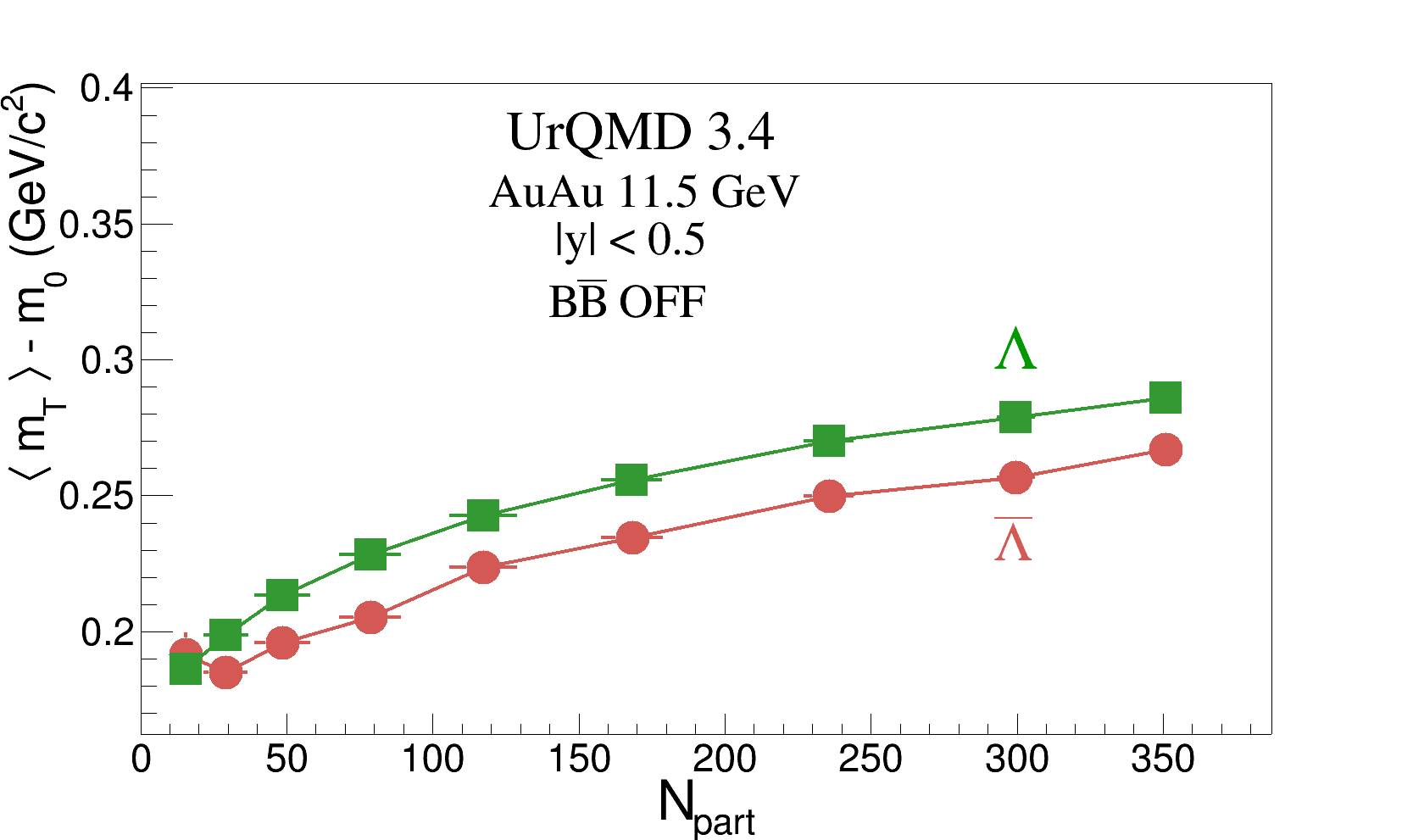}
	\caption{\label{mt_centrality_bboff} $\langle m_{T} \rangle$- m$_{0}$ for ${\Lambda}$ and $\bar{\Lambda}$ w.r.to N$_{part}$ at $\sqrt{s}_{NN}$ = 11.5 GeV, AuAu collision, mid rapidity $\mid y\mid$ $<$ 0.5 in UrQMD without including $\mathrm{B\bar{B}}$ annihilation.}
\end{figure}

It is interesting to note that UrQMD calculations without incorporating $\mathrm{B\bar{B}}$ annihilation also show a difference in the magnitudes of $\langle m_{T} \rangle$- m$_{0}$ for ${\Lambda}$ and $\bar{\Lambda}$ but with an opposite trend i.e, $\langle m_{T} \rangle$- m$_{0}$ of ${\Lambda}$ is systematically higher than $\bar{\Lambda}$. In absence of $\mathrm{B\bar{B}}$ annihilation, this split in $\langle m_{T} \rangle$- m$_{0}$ can be explained taking into account different energy thresholds of ${\Lambda}$ and $\bar{\Lambda}$ production in hadronic interaction channels. At lower $\sqrt{s}$ where baryon density is high, ${\Lambda}$s are dominantly produced in associated production channel N + N $\rightarrow$ ${\Lambda}$ + K$^{+}$ + N, which has an energy threshold of $\sim$ 700 MeV, whereas, $\bar{\Lambda}$ production is dominated by ${\Lambda}$-$\bar{\Lambda}$ pair production channel in N + N $\rightarrow$ ${\Lambda}$ + $\bar{\Lambda}$ + N + N
that has an energy threshold of $\sim$ 2200 MeV. Therefore, the energy, in excess to threshold energy, available to impart kinetic energy to the produced particles in ${\Lambda}$-$\bar{\Lambda}$ pair-production channel is small. Thus, produced $\bar{\Lambda}$s mostly have smaller transverse momenta and hence lower values of average transverse mass than ${\Lambda}$s.

We further study the interplay between $\mathrm{B\bar{B}}$ annihilation and threshold energy for producing ${\Lambda}$ and $\bar{\Lambda}$ by calculating $\langle m_{T} \rangle$- m$_{0}$ for ${\Lambda}$ and $\bar{\Lambda}$ as a function of collision centrality represented by the number of participating nucleons (N$_{part}$). Fig.~\ref{mt_centrality} shows the $\langle m_{T} \rangle$- m$_{0}$ for ${\Lambda}$ and $\bar{\Lambda}$ as a function of N$_{part}$ as obtained from the STAR data for Au+Au collisions at $\sqrt{s} =$ 11.5 GeV (left)~\cite{STAR_PRC_2020} and compared it with UrQMD calculations including $\mathrm{B\bar{B}}$ annihilation at the same $\sqrt{s}$ (right). Both data and UrQMD model exhibit similar behaviour in $\langle m_{T} \rangle$- m$_{0}$ for ${\Lambda}$ and $\bar{\Lambda}$ as a function of N$_{part}$. When the system size is small or at low N$_{part}$ ( N$_{part} <$ $\sim$~100), the so called ``threshold effect'' is dominant as the baryon density is low, as discussed earlier. This results in higher values of $\langle m_{T} \rangle$- m$_{0}$ for ${\Lambda}$ than $\bar{\Lambda}$ and then followed by a switch-over for N$_{part} >$ 100, where the trend get reversed i.e, $\langle m_{T} \rangle$- m$_{0}$ for $\bar{\Lambda}$ is greater than  ${\Lambda}$. This happens because as the system size increases, baryon stopping becomes higher which causes more baryons to accumulate at mid-rapidity. Thus, the effects of $\mathrm{B\bar{B}}$ annihilation start to become more significant causing low p$_{T}$ $\bar{\Lambda}$s to annihilate more than ${\Lambda}$s resulting in hardening of $\bar{\Lambda}$s p$_{T}$ spectra, as discussed earlier. As can be seen from Fig.~\ref{mt_centrality_bboff}, when $\mathrm{B\bar{B}}$ annihilation is not included in UrQMD, $\langle m_{T} \rangle$- m$_{0}$ for ${\Lambda}$ is higher than $\bar{\Lambda}$ at all N$_{part}$. This essentially suggests that in central collisions at lower $\sqrt{s}$ effects of $\mathrm{B\bar{B}}$ annihilation can not be ignored particularly while considering any phenomenon that involves yields or spectra of baryons and anti-baryons in the final state.

\begin{figure}[htbp]
		\centering
	\includegraphics[width=84mm,height=77mm]{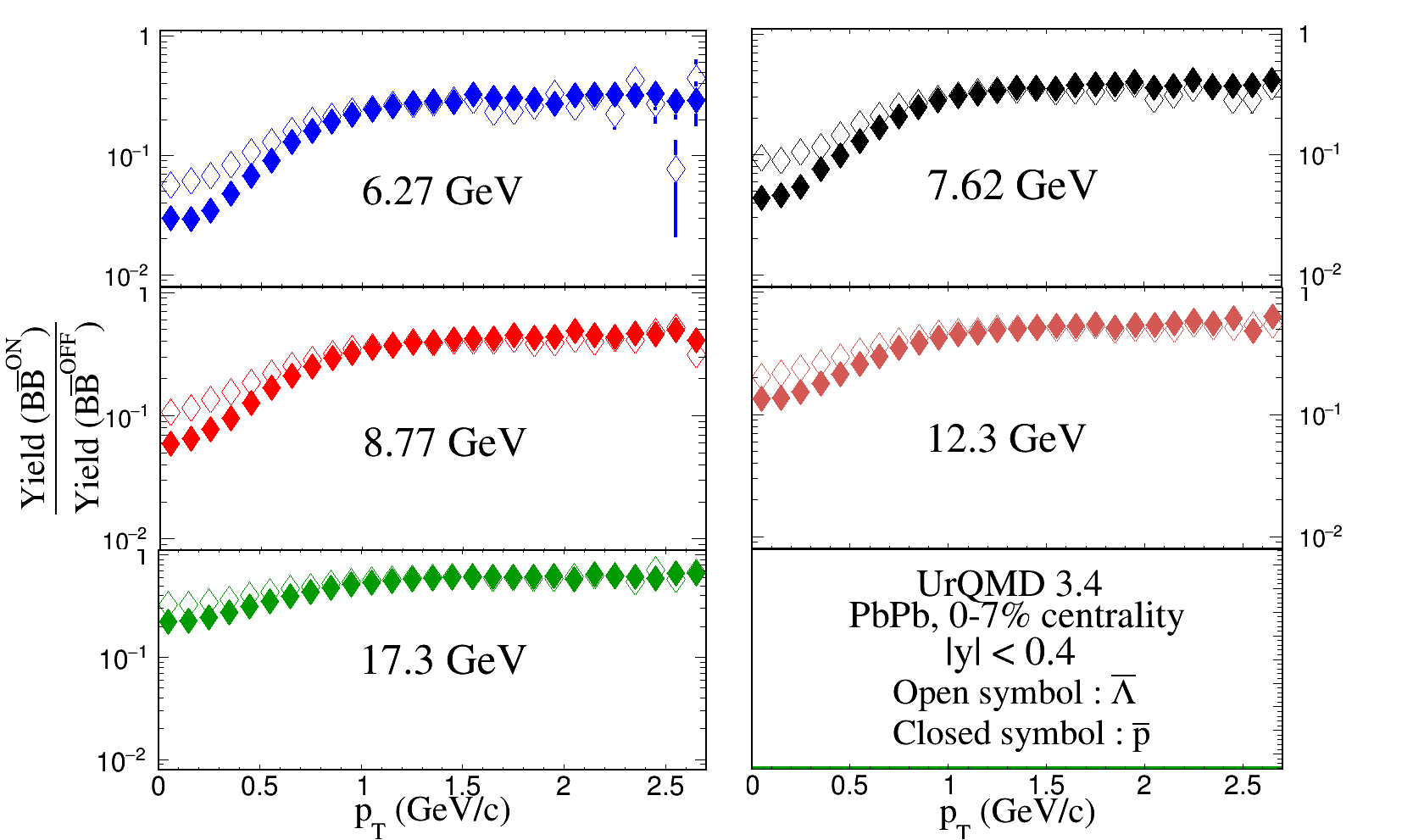}
	\caption{\label{pt_ratio_energy} The p$_{T}$ dependence of ratios of yields of $\bar{\Lambda}$'s (open marker) and $\bar{p}$'s (closed marker), with and without incorporating $\mathrm{B\bar{B}}$ annihilation calculated from UrQMD
in central Pb+Pb collisions at $\sqrt{s_{NN}} =$ 6.27 to 17.3 GeV. }
\end{figure}

\begin{figure}[htbp]
		\centering
	\includegraphics[width=84mm,height=77mm]{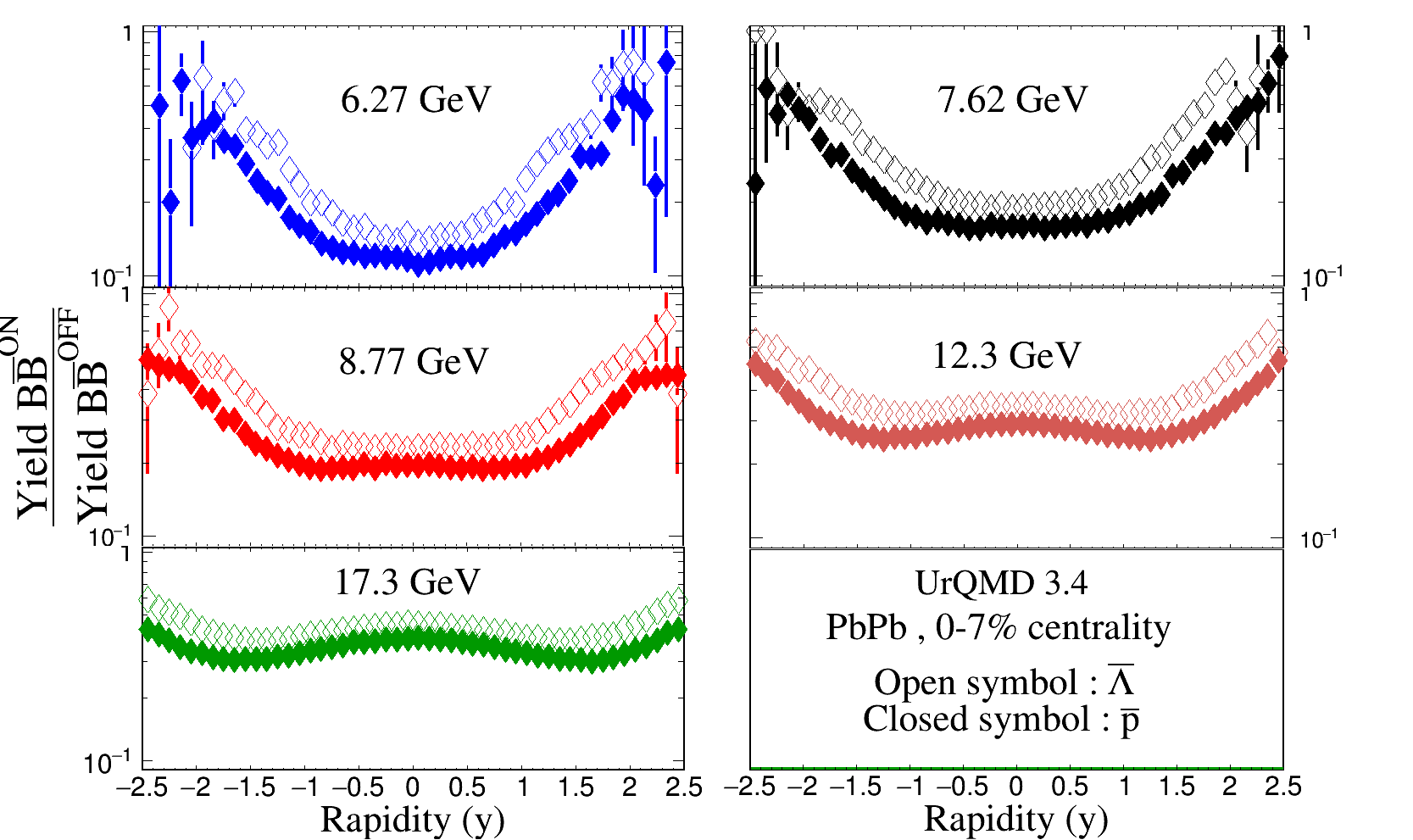}
	\caption{\label{y_ratio_energy} The rapidity dependence of ratios of yields of $\bar{\Lambda}$'s (open marker) and $\bar{p}$'s (closed marker), with and without incorporating $\mathrm{B\bar{B}}$ annihilation calculated from UrQMD in central Pb+Pb collisions at $\sqrt{s_{NN}} =$ 6.27 to 17.3 GeV. }
\end{figure}

\begin{figure}[htbp]
		\centering
	\includegraphics[width=80mm,height=55mm]{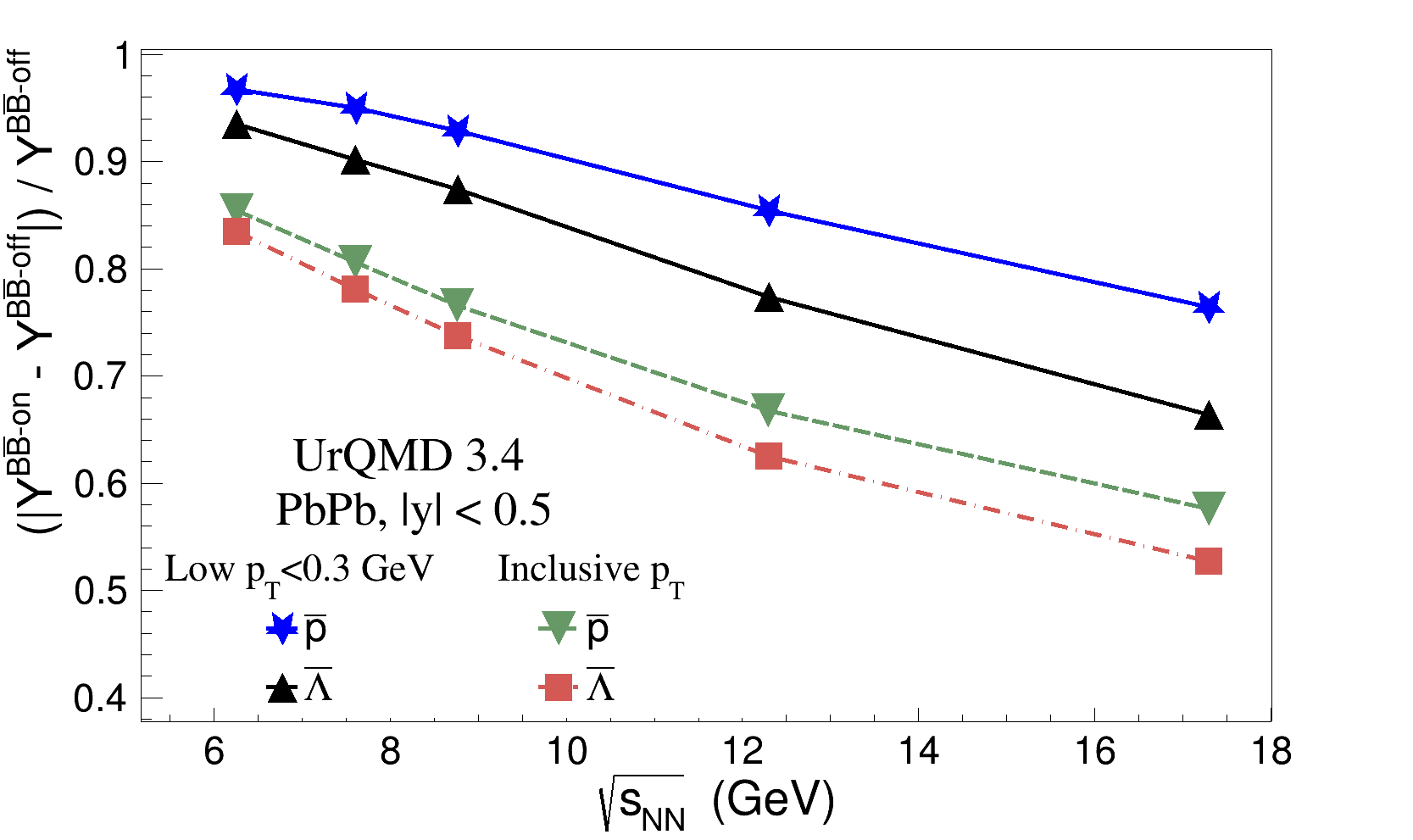}
	\caption{\label{annh_frac}Annihilation fraction of $\bar{p}$ and $\bar{\Lambda}$ as a function of $\sqrt{s}_{NN}$ for low and inclusive p$_{T}$ from UrQMD}
\end{figure}

Having seen that $\mathrm{B\bar{B}}$ annihilation can significantly impact final spectra of baryons and anti-baryons, we now proceed to study the effect of $\mathrm{B\bar{B}}$ annihilation on $\bar{\Lambda}$/$\bar{p}$ ratios. Before that first we investigate how $\mathrm{B\bar{B}}$ annihilation modify yields of $\bar{\Lambda}$ and $\bar{p}$ as a function of p$_{T}$ and rapidity ($\textit{y}$) at different $\sqrt{s}$. In Fig.~\ref{pt_ratio_energy}, we plot the ratio of yields  for $\bar{\Lambda}$ (open symbol) and $\bar{p}$ (closed symbol) as a function of p$_{T}$ with and without incorporating $\mathrm{B\bar{B}}$ annihilation in UrQMD. This plot suggests that the effect of annihilation is maximum at low-p$_{T}$ and there is a particle species dependence i.e, $\bar{p}$ is annihilated more than $\bar{\Lambda}$ thus, final yields of $\bar{p}$ are more suppressed than $\bar{\Lambda}$. For p$_{T} >$  1 GeV the effect of $\mathrm{B\bar{B}}$ annihilation on both $\bar{\Lambda}$ and $\bar{p}$ appear to be same. The overall nature of the p$_{T}$ dependent modification of $\bar{\Lambda}$ and $\bar{p}$ yields because of $\mathrm{B\bar{B}}$ annihilation appear to be similar at all $\sqrt{s}$ studied here except, the magnitude of suppression increases with decreasing $\sqrt{s}$, implying more annihilation of $\bar{\Lambda}$ and $\bar{p}$ at lower $\sqrt{s}$. Similarly, the effect of $\mathrm{B\bar{B}}$ annihilation on rapidity dependence of $\bar{\Lambda}$ and $\bar{p}$ yields are studied and plotted in Fig.~\ref{y_ratio_energy}. 
The annihilation effect is most significant at mid-rapidity in low $\sqrt{s}$ for both $\bar{\Lambda}$ and $\bar{p}$, characterized by a large suppression of ratio of yields obtained from model calculations with and without $\mathrm{B\bar{B}}$ annihilation. As energy increases, the overall magnitude of suppression start to decrease however, the maximum suppression now occurs at forward rapidity instead of mid-rapidity. This happens because with increasing energy baryon stopping is less and the high baryon density region gradually move from mid to forward rapidity. As a result, the effect of $\mathrm{B\bar{B}}$ annihilation at higher $\sqrt{s}$ becomes more significant at forward rapidity. Next we summarize the effect of $\mathrm{B\bar{B}}$ annihilation on $\bar{\Lambda}$ and $\bar{p}$ yields as a function of $\sqrt{s}$ for low and inclusive p$_{T}$ in Fig.~\ref{annh_frac}. In this figure we quantify the magnitude of annihilation interms of a quantity called annihilation fraction which is defined as relative difference in yields with and without $\mathrm{B\bar{B}}$ annihilation (Y$^\mathrm{B\bar{B}-on}$ - Y$^\mathrm{B\bar{B}-off}$)  scaled by the yield obtained without incorporating $\mathrm{B\bar{B}}$ annihilation.

From Fig.~\ref{annh_frac} it is clear that at low $\sqrt{s}$, say at 6.27 GeV, $>$ 95\% of the initially produced low-p$_{T}$ $\bar{p}$ are annihilated whereas, for $\bar{\Lambda}$ the annihilation fraction is about 92 to 94\%. For inclusive p$_{T}$, annihilation fraction is below 90\%  for both the particles. As the energy increases annihilation fraction drops, i.e lesser number of initially produced baryons are annihilated. It is seen that for $\bar{\Lambda}$, annihilation fraction as a function $\sqrt{s}$ drops faster than $\bar{p}$ for both inclusive and low-p$_{T}$ .

\begin{figure}[htbp]
	\includegraphics[width=88mm,height=52mm]{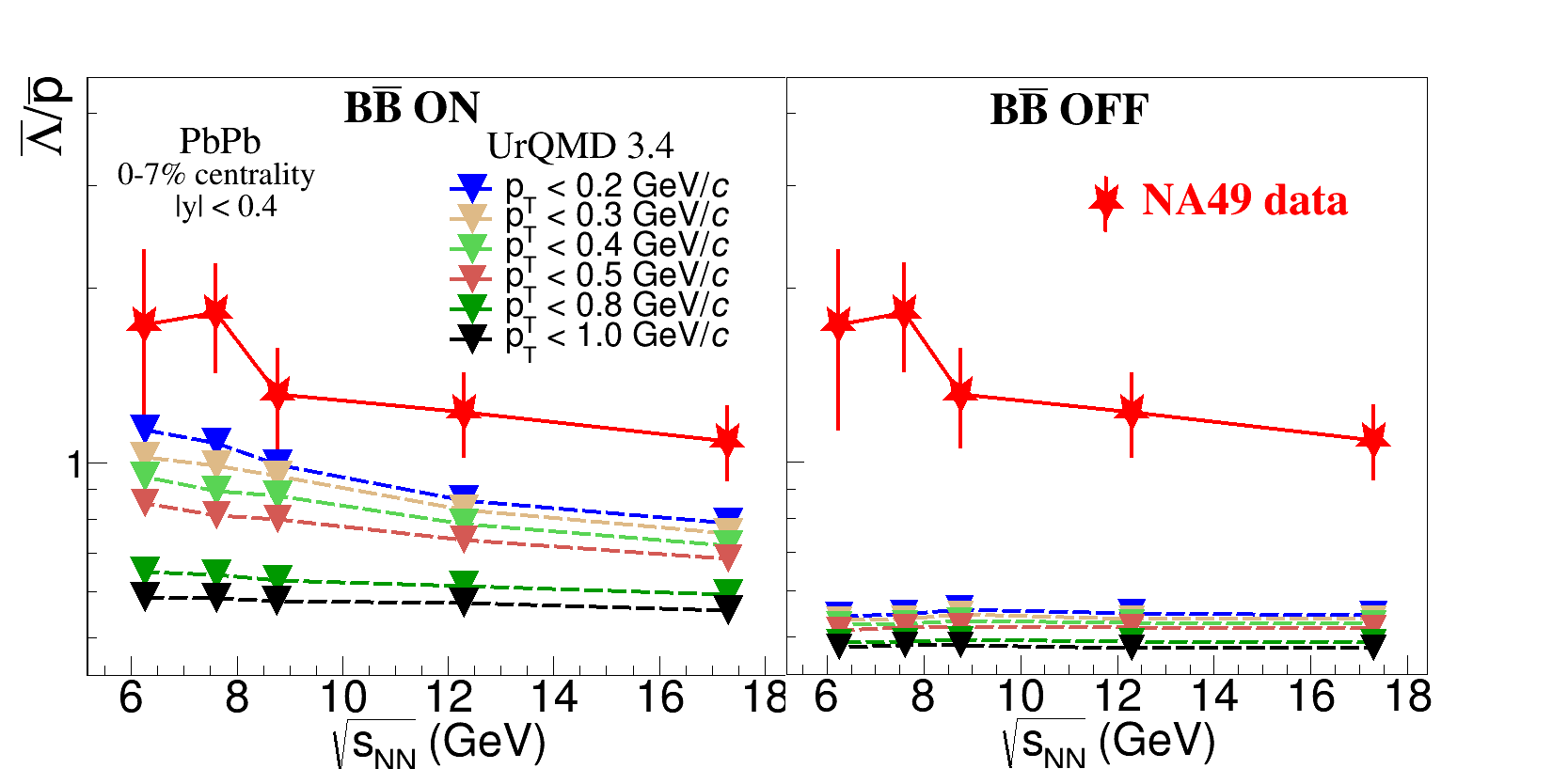}    
	\caption{\label{ratio_energy} The $\bar{\Lambda}$/$\bar{p}$ ratio  at mid-rapidity as a function of collision energy at $\sqrt{s_{NN}} =$ 6.27 to 17.3 GeV with (left) and without (right) $\mathrm{B\bar{B}}$-annihilation from UrQMD. Also shown the effect of different p$_{T}$ cut on the ratio. Model calculation is compared with NA49~\cite{CAlt_NA49_nucl_ex0512033} data (feed-down uncorrected) at same $\sqrt{s_{NN}}$. }
\end{figure}

\begin{figure}[htbp]
	\centering
	\includegraphics[width=78mm,height=55mm]{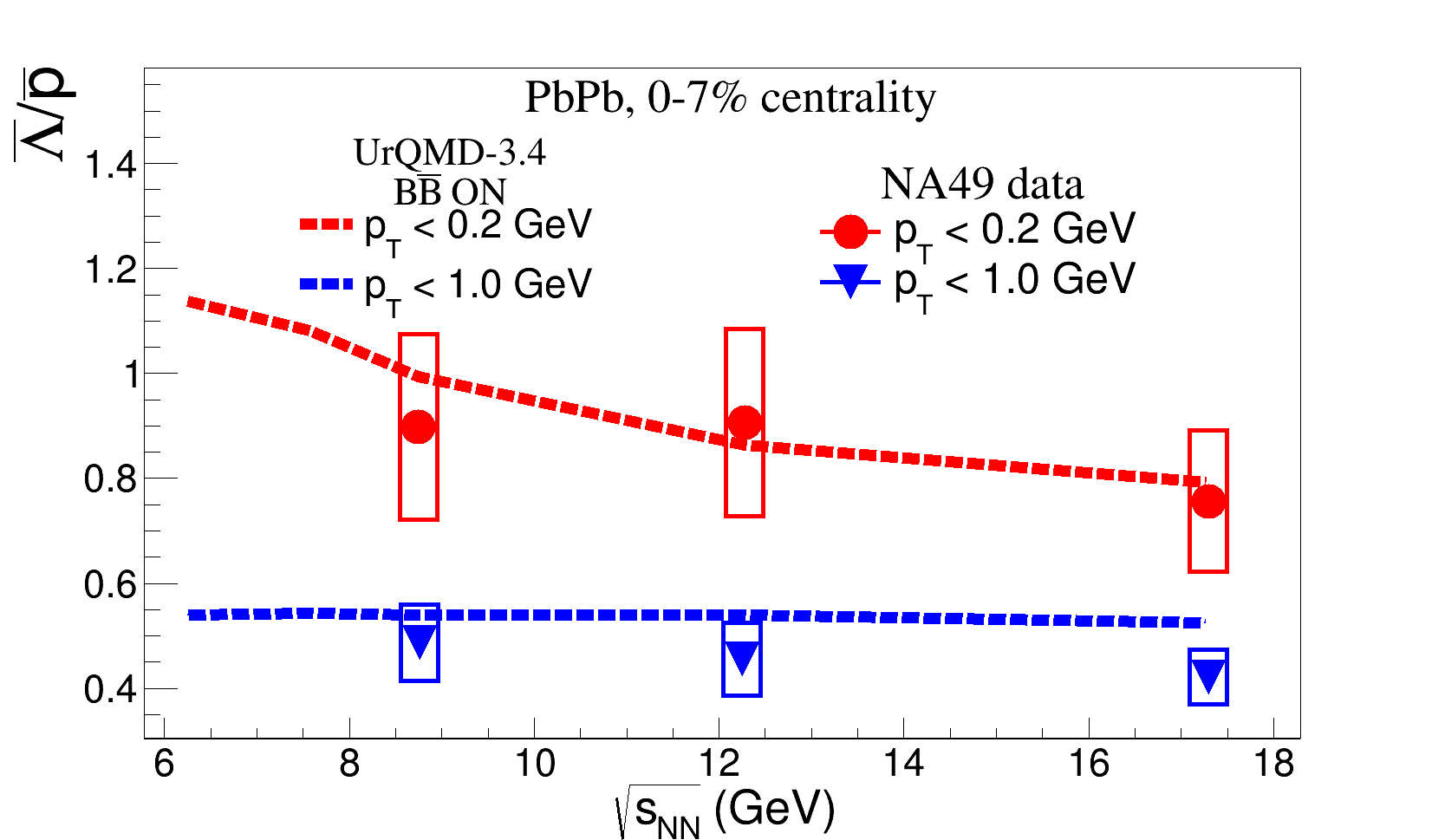}
	\caption{\label{ratio_energy_corrected} Dependence of $\bar{\Lambda}$/$\bar{p}$ ratios as a function of $\sqrt{s}_{NN}$ with $\mathrm{B\bar{B}}$ annihilation in UrQMD and compared it with feed-down corrected NA49 data points at low p$_{T}$ and inclusive p$_{T}$ for central (0-7$\%$) PbPb collisions.}
\end{figure}

Finally, we calculate $\sqrt{s}$ dependence of $\bar{\Lambda}$/$\bar{p}$ ratios with and
without $\mathrm{B\bar{B}}$ annihilation, as shown in Fig.~\ref{ratio_energy}, where NA49 data points are also included. We find that $\bar{\Lambda}$/$\bar{p}$ ratios are indeed sensitive to $\mathrm{B\bar{B}}$ annihilation and its impact depend strongly on the kinematic selection, in particular, to the choice of p$_{T}$-range. We observe that in low p$_{T}$ range ( p$_{T} < $ 0.2 GeV/c) $\bar{\Lambda}$/$\bar{p}$ ratios achieve maximum and it exceeds unity for the lowest collision energy. This trend for $\bar{\Lambda}$/$\bar{p}$ enhancement in the UrQMD model is qualitatively similar to data and somehow expected, because, annihilation cross-sections of $\bar{\Lambda}$ is less than $\bar{p}$ which result in more suppression of $\bar{p}$ yield in the final state compared to $\bar{\Lambda}$, leading to an enhancement in $\bar{\Lambda}$/$\bar{p}$ ratios. The effect is certainly more pronounced in low-p$_{T}$ as we have already seen in Fig.~\ref{pt_ratio_energy} that annihilation probability increases with decrease in p$_{T}$. It is also interesting to note when UrQMD calculations are done without incorporating $\mathrm{B\bar{B}}$-annihilation, irrespective of the choice of p$_{T}$ range no enhancement can be seen in $\bar{\Lambda}$/$\bar{p}$ ratios. Here we want to draw attention on a fact that the first report on $\bar{\Lambda}$/$\bar{p}$ ratios from NA49, as shown in Fig.~\ref{ratio_energy}~\cite{CAlt_NA49_nucl_ex0512033} are calculated from feed-down un-corrected $\bar{\Lambda}$ yields \cite{TAnt_NA49_prl} and we speculate this could be a main reason for quantitative disagreement between data and model calculation. Therefore to verify our speculation we now calculate $\bar{\Lambda}$/$\bar{p}$ ratios from the feed-down corrected data. Since feed-down corrected yields are not directly available for different p$_{T}$ range, we extract feed-down corrected yields for both $\bar{\Lambda}$ and $\bar{p}$ using Blast-wave fit~\cite{blastwave} to feed-down corrected p$_{T}$ spectra ~\cite{CAlt_NA49_nucl_0804} and integrating it over desired p$_{T}$ range. Results for the feed-down corrected $\bar{\Lambda}$/$\bar{p}$ ratios are shown in Fig.~\ref{ratio_energy_corrected} for low and inclusive p$_{T}$ at $\sqrt{s}$ = 8.77, 12.3 and 17.3 GeV together with the results from UrQMD model calculations including $\mathrm{B\bar{B}}$ annihilation at same $\sqrt{s}$ and in same p$_{T}$-range.

\begin{figure}[htbp]
		\centering
	\includegraphics[width=78mm,height=55mm]{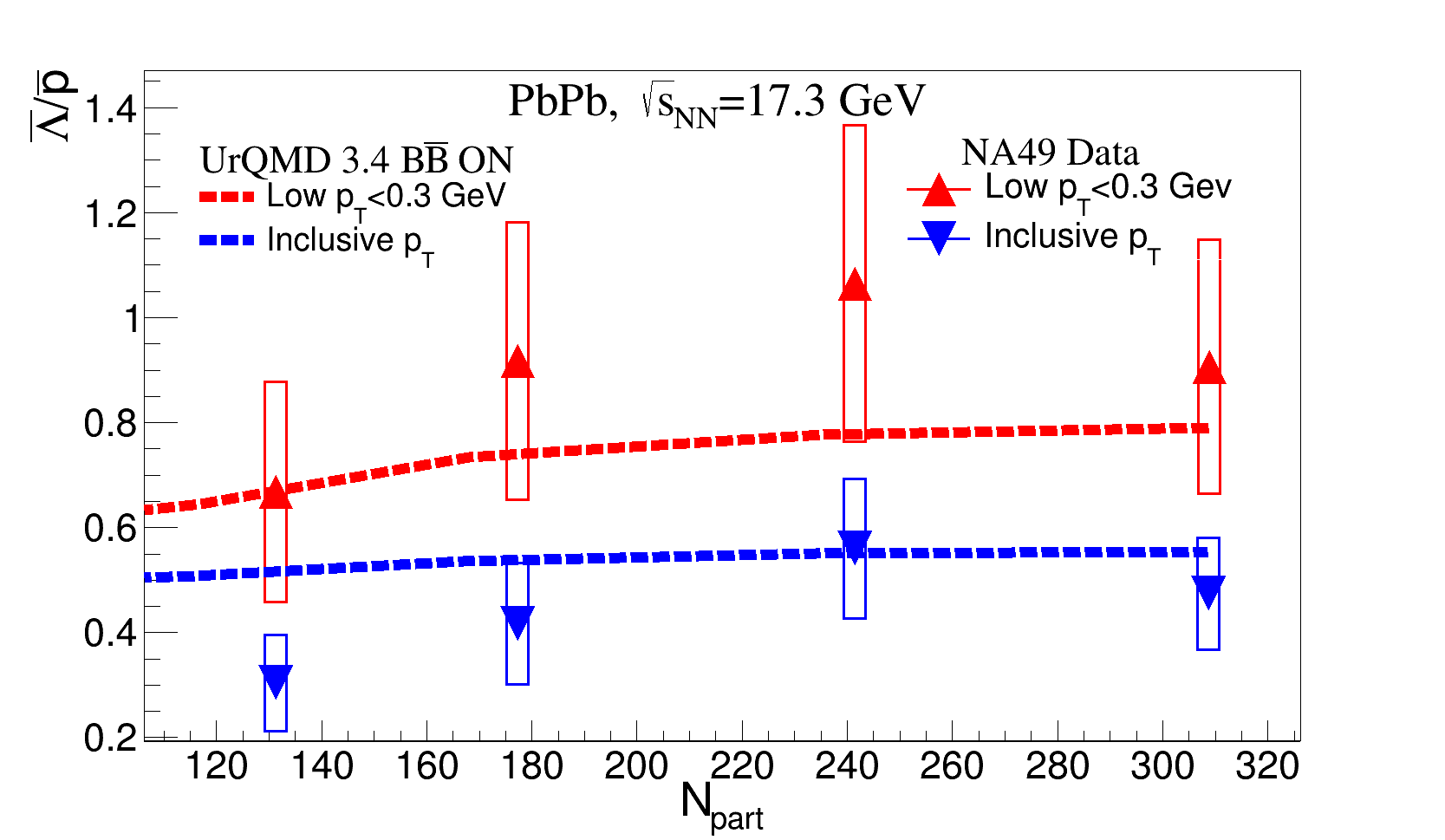}
	\caption{\label{ratio_centrality_corrected} $\bar{\Lambda}$/$\bar{p}$ ratios as a function of N$_{part}$ or centrality both from  UrQMD ($\mathrm{B\bar{B}}$ ON) and feed-down corrected NA49 data at $\sqrt{s}_{NN}$ = 17.3 GeV at low p$_{T}$ and inclusive p$_{T}$ }
\end{figure}

After the feed-down correction, $\bar{\Lambda}$/$\bar{p}$ ratios in data and model calculations show very good agreement, although with large uncertainty on data. It is to be noted that the enhancement in $\bar{\Lambda}$/$\bar{p}$ ratios both in data and model calculation is only observed at low-p$_{T}$ but for inclusive p$_{T}$  ratios are almost flat with $\sqrt{s}$. Therefore, it may be inferred that the observed enhancement in $\bar{\Lambda}$/$\bar{p}$ ratios in data could be an effect of $\mathrm{B\bar{B}}$ annihilation. We further study $\bar{\Lambda}$/$\bar{p}$ ratios as function of centrality both from data and UrQMD at $\sqrt{s}$ = 17.3 GeV. The ratios in data  are extracted from feed-down corrected yields using same procedure mentioned above and compared with UrQMD calculation with $\mathrm{B\bar{B}}$ annihilation in Fig.~\ref{ratio_centrality_corrected}. The above comparison between data and model indicate that
for inclusive p$_{T}$, $\bar{\Lambda}$/$\bar{p}$ ratios remain nearly flat as a function of centrality for both in data and model. For low-p$_{T}$, ratio increases from peripheral to central collision in UrQMD as well as in data although, uncertainties on data points are quite large. This systematic agreement between feed-down corrected data and UrQMD model including $\mathrm{B\bar{B}}$ annihilation further suggests that there is a strong impact of $\mathrm{B\bar{B}}$ annihilation on $\bar{\Lambda}$/$\bar{p}$ ratios and the enhancement in $\bar{\Lambda}$/$\bar{p}$ ratios may not be a necessary indication for strangeness enhancement in data.

\section{Summary}\label{sec4}

$\bar{\Lambda}$/$\bar{p}$ ratios have been measured at RHIC AGS and CERN SPS as a probe of strangeness enhancement. A large enhancement in the ratios were reported around 6-8 GeV collision energy, apparently consistent with the expectation of strangeness enhancement and hence to the onset of partonic deconfinement. However, at lower $\sqrt{s}$, where baryon density is large, hadronic interactions like $\mathrm{B\bar{B}}$ annihilation can significantly influence the individual spectra and yields of $\bar{\Lambda}$ and $\bar{p}$. In this work, we therefore studied whether $\mathrm{B\bar{B}}$ annihilation has any role in enhancement of $\bar{\Lambda}$/$\bar{p}$ ratios. To do so, we use UrQMD hadronic transport model taking into account $\mathrm{B\bar{B}}$ annihilation in the final state.

As a first step we established that $\mathrm{B\bar{B}}$ annihilation effects are significant at the SPS energy range by studying its effect on the average transverse mass spectra, $\langle m_{T} \rangle$- m$_{0}$ for $\Lambda$ and $\bar{\Lambda}$ as a function of $\sqrt{s}$ and compared it with NA49 and STAR data. It is seen that despite the same mass of $\Lambda$ and $\bar{\Lambda}$, the $\langle m_{T} \rangle$- m$_{0}$ for $\Lambda$ and $\bar{\Lambda}$ are different in magnitude, in particular $\langle m_{T} \rangle$- m$_{0}$ for $\bar{\Lambda}$ is higher than $\Lambda$, and the difference is large at low $\sqrt{s}$ and gradually reduces with increasing $\sqrt{s}$. UrQMD can qualitatively reproduce the trend in $\langle m_{T} \rangle$- m$_{0}$ when $\mathrm{B\bar{B}}$ annihilation is included. As a function of centrality, $\langle m_{T} \rangle$- m$_{0}$ for $\Lambda$ and $\bar{\Lambda}$ exhibits an interesting feature, at small N$_{part}$  $\langle m_{T} \rangle$- m$_{0}$ for $\Lambda$ is higher than $\bar{\Lambda}$ and then the trend becomes opposite i.e,  $\bar{\Lambda}$ has higher $\langle m_{T} \rangle$- m$_{0}$ than $\Lambda$ for N$_{part} >$ 100. This "switch over" can also be understood as a consequence of $\mathrm{B\bar{B}}$ annihilation because in UrQMD calculation without $\mathrm{B\bar{B}}$ annihilation, $\langle m_{T} \rangle$- m$_{0}$ for $\Lambda$ is systematically higher than $\bar{\Lambda}$ at all N$_{part}$.

Finally we studied the effect of $\mathrm{B\bar{B}}$ annihilation on $\bar{\Lambda}$/$\bar{p}$ ratios by comparing NA49 data with UrQMD calculations with $\mathrm{B\bar{B}}$ annihilation. In this case UrQMD could reproduce the qualitative trend in $\bar{\Lambda}$/$\bar{p}$ enhancement in data as a function $\sqrt{s}$. Subsequently we realize that $\bar{\Lambda}$ yields used for calculating the ratios in data are not feed-down corrected. Thus, we extract yields for $\bar{\Lambda}$ and $\bar{p}$ from the feed-down corrected spectra using a Blast-Wave fit to the corrected spectra and then use corrected yields to compute the $\bar{\Lambda}$/$\bar{p}$ ratios for low and inclusive p$_{T}$. The feed-down corrected $\bar{\Lambda}$/$\bar{p}$ ratios show enhancement as a function of $\sqrt{s}$ and centrality only for low-$p_{T}$ whereas, for inclusive $p_{T}$ no such enhancement is observed. The systematic of feed-down corrected $\bar{\Lambda}$/$\bar{p}$ ratios are seen to be in good agreement with UrQMD calculations including $\mathrm{B\bar{B}}$ annihilation. This suggests that the observed enhancement $\bar{\Lambda}$/$\bar{p}$ ratios is not necessarily related to strangeness enhancement but can also be interpreted as $\mathrm{B\bar{B}}$ annihilation effect. However, these findings can be concluded firmly from the measurements at the future facilities like FAIR where several orders of magnitude more data are foreseen to be collected.

\bmhead{Acknowledgments}
This research has used resources of grid computing facility at Variable Energy Cyclotron Centre (VECC), Kolkata. Authors would like to acknowledge discussions with Prof. Steffen A. Bass on baryon-anti-baryon annihilation in UrQMD.




\end{document}